# Simulation of centrifugal separation of particles with equal sedimentation velocities

Yurika Miyake[a], Natsuki Asami[b], Ken-ichi Amano[a,*]

[a] *Faculty of Agriculture, Meijo University, Nagoya 468-8502, Japan*

[b] *Graduate School of Agriculture, Meijo University, Nagoya 468-8502, Japan*

[*] Corresponding author. *E-mail address:* amanok@meijo-u.ac.jp (K. Amano).

## Abstract

Conventional centrifugal sedimentation cannot readily separate particles with the same sedimentation velocities. Density gradient centrifugation, although highly selective, exposes delicate particles such as drug delivery system (DDS) nanoparticles and exosomes to high osmotic pressure that can cause shrinkage, rupture, or expansion. To separate particles with equal or similar sedimentation velocities while preserving their properties, we propose two methods that exploit thermal fluctuations (Brownian motion): the centrifugal Brownian ratchet method, which uses asymmetric structures, and the centrifugal wavy-channel method, which uses geometric barriers. Two-dimensional Brownian dynamics simulations combining Stokes–Einstein diffusion with centrifugal sedimentation were performed, and particle transport was analyzed using the Péclet number. In the ratchet method, particles with identical sedimentation velocities but different Péclet numbers showed statistically distinct mean arrival positions and times. In the wavy-channel method, periodic entropic barriers interacting with differences in transport properties produced arrival-time differences, achieving temporal separation even for particles with identical sedimentation velocities or Péclet numbers. Both methods separate particles without relying on sedimentation-velocity differences and are expected to preserve particle properties. Because they neither depend on electrical properties nor require a change of medium, they may also be applicable to particles in liquid metals, oils, organic solvents, and ionic liquids.

## 1. Introduction

Centrifugal sedimentation and density gradient centrifugation are widely used as conventional centrifugation-based particle separation techniques. Centrifugal sedimentation separates particles on the basis of differences in sedimentation velocity arising from differences in particle size and density. However, particles with the same sedimentation velocity are difficult to separate by conventional centrifugation (Plüisch et al., 2019; Yakubovich et al., 2022; Smart and Lattuada, 2026). For example, a small particle with a high density and a large particle with a low density can have the same sedimentation velocity. Density gradient centrifugation, by contrast, offers high separation performance but requires a density gradient medium containing a high concentration of solute. The resulting high osmotic pressure may damage the structure and surface properties of delicate particles such as drug delivery system (DDS) nanoparticles and exosomes (Norouzi et al., 2017; Konoshenko et al., 2018; Yang et al., 2020). Conventional centrifugation-based separation techniques therefore face the challenge that particles with similar sedimentation velocities are difficult to separate while their properties are preserved.

To address this issue, separation methods that utilize transport properties other than sedimentation velocity can be considered. A Brownian ratchet is a non-equilibrium transport mechanism that generates unidirectional particle transport by exploiting Brownian motion driven by thermal fluctuations together with spatial asymmetry generated by ratchet structures (Astumian, 1997; Grimm et al., 2009). The concept originates from the ratchet mechanism proposed by Smoluchowski in 1912 and was subsequently developed through Feynman's thermodynamic considerations and applications to molecular motors (Reimann, 2002). In a Brownian ratchet, no net directed transport occurs at thermal equilibrium; the system must be driven out of equilibrium, for example by an external perturbation (Linke et al., 2005). Moreover, because the direction and magnitude of transport change with the external perturbation and the ratchet structure, appropriate design of these conditions is important. The effectiveness of Brownian ratchets for particle transport and separation depends on the appropriate design of the driving conditions and spatial asymmetry, suggesting their potential as an effective approach for small-particle separation (Xuan and Lee, 2014).

Separation using wavy channels, on the other hand, is based on the force-driven transport of Brownian particles across geometric confinement. This approach controls particle transport through the competition among the entropic barriers created by the geometric confinement of the channel, the external force, and Brownian motion. The concept was developed on the basis of Fick–Jacobs-type diffusion theory (Kalinay and Percus, 2006; Berezhkovskii et al., 2015) and was extended to particle transport in periodic channels subjected to an external force (Reguera and Rubí, 2001; Schmid et al., 2009; Reguera et al., 2012). Laachi et al. (2007) analyzed the mean velocity and effective diffusion coefficient of Brownian particles subjected to a constant external force in channels with periodically varying cross-sections, and evaluated the competition between the external force and Brownian motion using the Péclet number. Furthermore, because particles exhibit size-dependent transport properties, the potential of this approach for particle separation has also been shown (Riefler et al., 2010). However, transport properties depend strongly on the channel geometry and the external-force conditions, and the applicability of the Fick–Jacobs approximation is limited for steep channel geometries or strong external forces (Reguera et al., 2006; Burada et al., 2007). Transport in wavy channels is thus expected to find application in particle separation as a means of controlling the transport of small particles through the channel geometry and the external force (Burada et al., 2008).

Previous studies have demonstrated force-driven particle separation using periodic obstacle arrays under gravity or centrifugal acceleration, a technique known as deterministic lateral displacement (DLD) (Devendra and Drazer, 2012; Du and Drazer, 2016; Jiang et al., 2016). However, these studies focused on relatively large particles for which Brownian motion is negligible and achieved separation based on differences in particle size. In contrast, the present study focuses on relatively small particles whose transport is influenced by Brownian motion and investigates separation by exploiting differences in transport behavior arising from the interplay between Brownian motion and centrifugal force or from channel geometry. To the best of the authors' knowledge, no previous studies have investigated the separation of particles with different densities but identical sedimentation velocities using either a Brownian ratchet under centrifugal force or a wavy channel with periodically varying width under centrifugal force. These methods may provide alternative means of separating particles that are difficult to distinguish by conventional centrifugation. Furthermore, incorporating ratchet structures or wavy channels into a dedicated centrifuge tube and tube holder could enable these separation principles to be implemented using existing centrifuges. These considerations motivate a systematic investigation of the fundamental transport behavior and separation characteristics of the two methods, particularly for particles that are difficult to separate by conventional centrifugation.

Therefore, in this study, we propose particle separation methods using Brownian ratchets and wavy channels under centrifugal force, referred to as the centrifugal Brownian ratchet method and the centrifugal wavy-channel method, respectively. Their separation performance was compared for particles with different sedimentation velocities, identical sedimentation velocities, and identical Péclet numbers. In the centrifugal Brownian ratchet method, ratchets were arranged in a staggered pattern inside a centrifuge tube, and particles were introduced at the left end of the top surface. In the centrifugal wavy-channel method, particles were introduced into a wavy channel from above. Brownian dynamics simulations were performed for both methods to evaluate particle transport and separation behavior. For both methods, we focused on the Péclet number, which represents the competition between transport by the external force and Brownian diffusion, and examined separation that exploits differences in particle transport properties. This approach may enable particles with equal or similar sedimentation velocities, which are difficult to separate by conventional centrifugation.

The proposed methods may also be applicable to the separation of particles dispersed in liquid metals. Liquid metals have higher electrical conductivity than common liquid media such as water and oil, which make electrophoretic separation of particles difficult to apply. In addition, extracting the particles to be separated from a liquid metal and transferring them to another medium may alter their surface state and properties. Therefore, even when particles must be separated while remaining in the liquid metal, the proposed methods, which exploit particle transport properties without relying on electrical properties, are expected to provide a useful separation principle. The separation without changing the liquid conditions may likewise be possible in oils, organic solvents, ionic liquids, and similar media.

The remainder of this paper is organized as follows. Section 2 describes the theory of the Brownian dynamics simulations and the system setups for the centrifugal Brownian ratchet method and the centrifugal wavy-channel method. Section 3 examines the separation performance of each method for particles with different or similar sedimentation velocities, as well as for particles with equal Péclet numbers. Section 4 (Conclusions) compares the separation performance of the two methods and discusses their characteristics.

## 2. Theory

### *2.1. Brownian dynamics simulation*

The simulations were performed with a time step $\Delta t$ = $1.0 \times 10^{-5}$ s and a total number of computational steps $n_{\mathrm{step}}$ = $5 \times 10^{8}$ for the centrifugal Brownian ratchet method, and a time step $\Delta t$ = $6.0 \times 10^{-5}$ s and a total number of computational steps $n_{\mathrm{step}}$ = $1 \times 10^{10}$ for the centrifugal wavy-channel method. The time step was set to a value small enough that a particle could not pass through a ratchet obstacle in a single step, and the total number of steps was set large enough for all particles to reach the arrival plane.

The random motion of particles due to thermal fluctuations was modeled as Brownian motion. Brownian motion is the phenomenon in which a particle, continually struck by the thermally moving molecules of the surrounding liquid, exhibits nearly irregular motion. This motion becomes more pronounced as the particle size decreases.

With the Boltzmann constant $k_{\mathrm{B}}$, the absolute temperature $T$, the fluid viscosity $\eta$, and the particle radius $r$, the diffusion coefficient $D$ due to Brownian motion is given by the Stokes–Einstein relation:

$$D = \frac{k_B T}{6\pi\eta r}. \tag{1}$$

In this study, the random displacement at each time step was generated from this diffusion coefficient to reproduce Brownian motion. The displacement due to Brownian motion in one step is expressed as

$$\Delta l = \sqrt{2D\Delta t}\,\xi, \tag{2}$$

where $\xi$ is a random number following the standard normal distribution with a mean of 0 and a variance of 1. The mean displacement is therefore zero, and its magnitude varies randomly on the scale of $\sqrt{2D\Delta t}$. In the simulations, a random number $\xi$ was generated for each particle at each time step, and the corresponding displacement was applied to reproduce Brownian motion.

An important concept for understanding Brownian motion is the principle of detailed balance. Detailed balance is the principle that, at thermal equilibrium, the transitions occurring per unit time between any two states $i$ and $j$ balance each other, as expressed by $P_i W_{i\to j} = P_j W_{j\to i}$. Here, $P_i$ and $P_j$ are the probabilities that a particle is in states $i$ and $j$, respectively, and $W_{i\to j}$ and $W_{j\to i}$ are the transition probabilities from state $i$ to state $j$ and from state $j$ to state $i$, respectively (Zhang et al., 2012; Dixit and Dill, 2014). This relation means that the number of particles moving in one direction exactly balances the number of particles moving in the opposite direction. Consequently, under thermal equilibrium, no net flow in any particular direction arises when averaged over long times, even though particles are always moving locally. For example, for particles undergoing only Brownian motion in a left–right symmetric space, the probabilities of moving to the right and to the left are equal, so the mean displacement is zero. This is the essence of detailed balance.

In this study, in addition to these random displacements due to thermal fluctuations, directionality due to centrifugal force was introduced into the particle motion. Particle inertia was neglected, and the Stokes regime was assumed, in which the Stokes drag always balances the centrifugal force corresponding to the centrifugal acceleration $a_{\mathrm{c}}$. Accordingly, each particle was assigned the sedimentation velocity $v_{\mathrm{s}}$ in the centrifugal field obtained from Stokes' drag law:

$$v_{\mathrm{s}} = \frac{2r^2(\rho_{\mathrm{p}} - \rho_{\mathrm{f}})a_{\mathrm{c}}}{9\eta}, \qquad (3)$$

where $\rho_{\mathrm{p}}$ is the particle density and $\rho_{\mathrm{f}}$ is the fluid density. With this terminal velocity $v_{\mathrm{s}}$, the drift displacement due to centrifugal force during $\Delta t$ is expressed as $v_{\mathrm{s}}\Delta t$. The particle motion was therefore computed by superposing the random displacement due to Brownian motion and the drift displacement due to centrifugal force. That is, taking the direction in which the centrifugal force acts as the $y$ direction, the particle displacement at each time step was given as follows:

$$\Delta x = \sqrt{2D\Delta t}\,\xi_{\mathrm{x}}, \tag{4}$$

$$\Delta y = v_{\mathrm{s}}\Delta t + \sqrt{2D\Delta t}\,\xi_{\mathrm{y}}, \tag{5}$$

where $\xi_{\mathrm{x}}$ and $\xi_{\mathrm{y}}$ are independent standard normal random numbers. As shown in Eqs. (4) and (5), particles diffuse randomly through Brownian motion while, on average, being transported in a fixed direction by the centrifugal force.

To evaluate the relative strength of centrifugal drift and Brownian diffusion, the Péclet number was introduced. The Péclet number is a dimensionless number representing the relative contributions of transport by centrifugal and diffusive forces. In this study, the characteristic length $L$ was defined differently for the two methods. For the centrifugal Brownian ratchet method, $L$ was defined as the diagonal distance between adjacent lattice positions of the ratchet obstacles, calculated as $L = \sqrt{l_{\mathrm{x}}^2 + l_y^2} = 25.6\ \mu\mathrm{m}$, where $l_{\mathrm{x}}$ and $l_{\mathrm{y}}$ are the spacings between adjacent lattice positions in the $x$ and $y$ directions, respectively. Using the characteristic length $L$, the drift velocity due to centrifugal force $v_{\mathrm{s}}$, and the diffusion coefficient $D$, the Péclet number was defined from Eqs. (1) and (3) as

$$N_{\mathrm{P}} = \frac{|v_{\mathrm{s}}|L}{D} = \frac{4\pi r^3 |\rho_{\mathrm{p}} - \rho_{\mathrm{f}}| a_{\mathrm{c}} L}{3k_{\mathrm{B}}T}. \tag{6}$$

For the centrifugal wavy-channel method, $L$ was defined as the channel width, $L = 2W_{\mathrm{h}} = 40\ \mu\mathrm{m}$, where $W_{\mathrm{h}}$ is the channel half-width. To characterize the competition between transverse diffusion and drift along the channel, a modified Péclet number, $N_{\mathrm{P,w}}$, was defined as the ratio of the transverse diffusion time to the time required to travel one channel wavelength:

$$N_{\mathrm{P,w}} = \frac{t_{\mathrm{d}}}{t_{\mathrm{a}}} = \frac{\left[\dfrac{L^2}{2D}\right]}{\left[\dfrac{\lambda}{|v_{\mathrm{s}}|}\right]} = \frac{L^2|v_{\mathrm{s}}|}{2D\lambda} = \frac{2\pi L^2 r^3 |\rho_{\mathrm{p}} - \rho_{\mathrm{f}}| a_{\mathrm{c}}}{3k_{\mathrm{B}}T\lambda}, \tag{7}$$

where $t_d$ is the characteristic time for diffusion across the full channel width, $t_a$ is the time required for drift over one channel wavelength, and $\lambda$ is the channel wavelength. For the centrifugal Brownian ratchet method, $N_P < 1$ indicates that diffusion is relatively dominant over drift, whereas $N_P > 1$ indicates that drift is relatively dominant. For the centrifugal wavy-channel method, $N_{P,w} < 1$ indicates that transverse diffusion is faster than drift over one wavelength, whereas $N_{P,w} > 1$ indicates the opposite. Increasing the centrifugal acceleration $a_c$ increases both Péclet numbers under otherwise fixed conditions.

Eqs. (6) and (7) also shows that the Péclet number depends on the particle size and mass density. For particles of the same density, a particle with a larger size increases the centrifugal drift velocity and decreases the diffusion coefficient, so that the Péclet number increases. In addition, for particles of the same size, a particle with a higher density increases the centrifugal drift velocity, while the diffusion coefficient remains unchanged, so that the Péclet number increases. In other words, the Péclet number increases with increasing particle size and particle density.

Accordingly, the particle transport properties in this study also depend on the particle size and particle density. The particle size affects both the strength of Brownian motion and the centrifugal drift, whereas the particle density mainly changes the magnitude of the centrifugal drift only. These differences give each particle a different Péclet number and change the relative contributions of centrifugal drift and Brownian diffusion.

Particle separation in this study exploits differences in the relative strength of Brownian diffusion and centrifugal drift that arise from different combinations of particle size and particle density. These differences can be characterized by the Péclet number, and particle separation is considered to become possible because particle behavior differs according to the Péclet number of each particle.

The solvent conditions in the simulations were a temperature of 298.15 K, a viscosity of 0.001 Pa·s, and a density of 1000 kg/m$^3$. Of the particles used for separation, the particles with smaller or equal sizes were designated as particles A, and those with larger or equal sizes as particles B. The particle radii are $r_A$ and $r_B$ (m), and the particle densities are $\rho_{pA}$ and $\rho_{pB}$ (kg/m$^3$), respectively. The numbers of particles A and B were denoted by $n_A$ and $n_B$, respectively. In the simulations, 2000 particles A ($n_A$ = 2000) and 2000 particles B ($n_B$ = 2000) were used. Particle–particle interactions and Faxén/Brenner effects were not considered, and each particle was treated as an independent Brownian particle. This assumption was made because dilute solution conditions were considered and because the fundamental effects of the designed ratchets and wavy channels were to be investigated.

### *2.2. System setup for the centrifugal Brownian ratchet method*

For the centrifugal Brownian ratchet method, a centrifuge tube containing ratchets with spatial asymmetry was designed (see Fig. 1). The computational domain was a centrifuge tube with a width from $x_{\min}$ = 0 to $x_{\max}$ = 1 mm and a depth from $y_{\min}$ = 0 to $y_{\max}$ = 1.1 mm, inside which asymmetrically shaped ratchet obstacles were arranged periodically. The system was designed on a relatively small scale to reduce the computational cost. Nevertheless, this scale allowed the fundamental effects of the ratchets to be evaluated. In an actual centrifuge tube, the characteristic dimensions are typically on the centimeter scale. If a larger-scale system were constructed, the radial variation in the direction of the centrifugal force with respect to the rotation axis would need to be considered.

A parallelogram, which has a large asymmetry and readily produces a rectification effect, was adopted as the ratchet shape. The ratchet width and height were $w$ = 8 μm and $h$ = 5 μm, respectively. Shape asymmetry was introduced by translating the upper side of the parallelogram horizontally by $x_s$ = 7 μm. In addition, the entire obstacle was rotated by an angle $\theta$ = 0.45 rad to prevent particles from being trapped on horizontal surfaces and to induce a rectification effect.

The ratchets were placed at a distance $d_w$ = 5 μm from the tube walls and repeated periodically with a horizontal spacing $l_x$ = 20 μm and a vertical spacing $l_y$ = 16 μm. The even-numbered and odd-numbered rows were shifted horizontally relative to each other by $\alpha$ = 0.5 times $l_x$ to form a staggered arrangement.

Particles A and B were initially placed at random within the leftmost 1/16 of the tube width at the upper end of the $y$ axis. A constant centrifugal acceleration $a_c$ (m/s$^2$) was applied to the particles vertically downward (in the $+y$ direction). The particles were transported downward by the centrifugal force while undergoing Brownian motion, and the simulation was terminated when all particles had reached the arrival plane $y_a$ = 1 mm (slightly above $y_{\max}$).

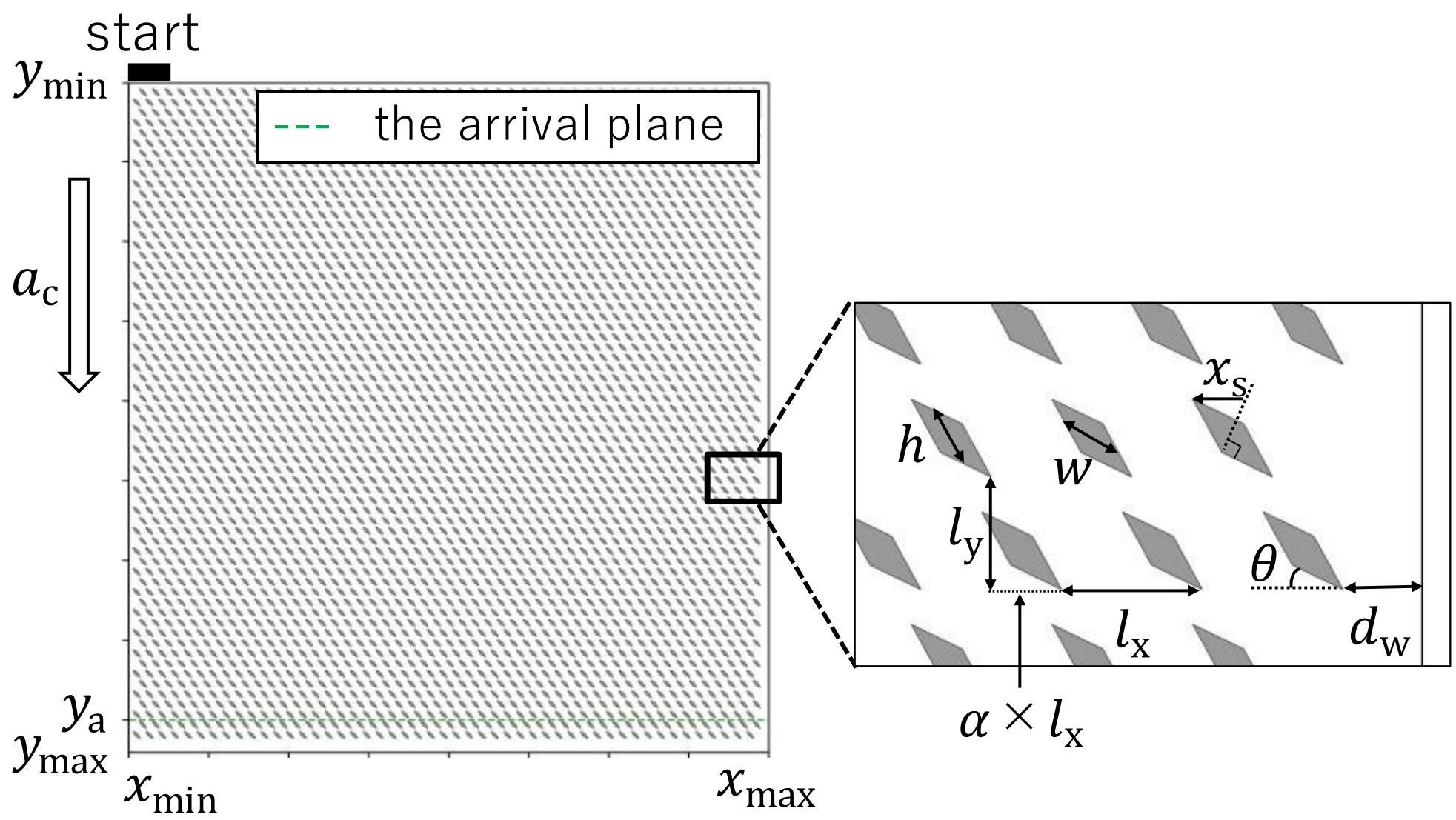

**Fig. 1.** Schematic of the simulated system. The system was designed with the following values: $x_{\mathrm{min}}$ = 0, $x_{\mathrm{max}}$ = 1 mm, $y_{\mathrm{min}}$ = 0, $y_{\mathrm{max}}$ = 1.1 mm, $y_{\mathrm{a}}$ = 1 mm, $w$ = 8 μm, $h$ = 5 μm, $x_{\mathrm{s}}$ = 7 μm, $\theta$ = 0.45 rad, $l_x$ = 20 μm, $l_{\mathrm{y}}$ = 16 μm, $d_{\mathrm{w}}$ = 5 μm, and $\alpha$ = 0.5 (staggered arrangement).

Collisions with walls were detected by treating each particle as a rigid sphere of finite radius. The ratchet obstacles were expanded by the radius of each particle, and overlap with an obstacle was determined using these effective obstacle regions. At each time step, the particle position was first updated according to the Brownian displacement and the centrifugal drift, with the displacements in the $x$ and $y$ directions calculated using Eqs. (4) and (5), respectively. If the resulting position overlapped an effective obstacle, the particle was moved only in the $x$ direction by retaining the updated $x$ coordinate and restoring the $y$ coordinate to its pre-displacement value. If this position also overlapped an effective obstacle, the particle was instead moved only in the $y$ direction by restoring the $x$ coordinate to its value before the displacement and retaining the updated $y$ coordinate. If both of these positions overlapped an effective obstacle, the particle position was restored to its original position for that time step. Thus, the particle was allowed to move in one coordinate direction when possible while preventing overlap with the ratchet obstacles. Elastic reflection was applied separately at the outer walls.

In this study, rectification of Brownian motion was induced by placing spatially asymmetric ratchet structures under external driving by centrifugal force. At thermal equilibrium, detailed balance holds, so Brownian motion alone produces no net unidirectional transport. When external driving by centrifugal force breaks detailed balance and maintains the system out of equilibrium, however, particle motion due to thermal fluctuations can be rectified by the spatially asymmetric ratchet structures (Hänggi et al., 2005). When a particle collides with a ratchet obstacle, the asymmetric shape of the obstacle restricts the direction of motion and induces sliding along the obstacle surface. The geometric asymmetry of the obstacle then causes unequal rightward and leftward motions, so that the originally random motion due to thermal fluctuations is rectified in a specific direction (Huang et al., 2003).

The particle transport behavior induced by the ratchets differs according to the Péclet number of each particle (Kowalik and Bishop, 2016). For particles with a small Péclet number, for which diffusion is relatively dominant, random displacements readily occur even during sliding along the ratchet surface, producing motion not only in the rectified direction but also in the opposite directions. As a result, motion along the obstacle surface is disturbed, and the directional selectivity of the ratchet is less likely to be reflected. For particles with a large Péclet number, centrifugal drift is relatively dominant, the influence of random motion becomes relatively small, and the rectification effect of the ratchets becomes more pronounced.

Consequently, particles whose Péclet numbers differ because of differences in particle size and density interact differently with the ratchet obstacles and exhibit different transport behavior. Because these differences appear as differences in the mean transport and the arrival position, even particles with equal or similar sedimentation velocities are considered to be separable in the lateral direction at the bottom of the system.

### *2.3. System setup for the centrifugal wavy-channel method*

In the centrifugal wavy-channel method, a centrifuge tube with a wavy channel was used (see Fig. 2(a)). For comparison, a centrifuge tube with a straight channel was also used (see Fig. 2(b)).

The channel was modeled as a two-dimensional wavy channel whose left and right walls vary sinusoidally. The channel half-width was $W_{\mathrm{h}}$ = 20 μm, the amplitude of the wavy walls $A$ = 15 μm, the wavelength $\lambda$ = 15 μm, and the channel extended from $y_{\mathrm{min}}$ = 0 to $y_{\mathrm{max}}$ = 10.1 mm. The left and right walls were varied in antiphase, so that the channel width periodically narrows and widens. For comparison, a straight channel with width of $2W_{\mathrm{h}}$ was also designed. Particles A and B were initially placed randomly at the upper end of the $y$ axis.

A constant centrifugal acceleration $a_c$ (m/s$^2$) was applied to the particles vertically downward (in the $+y$ direction). The particles were transported downward by the centrifugal force while undergoing Brownian motion, and the time at which a particle reached the arrival plane $y_a$ = 10 mm (slightly above the channel end $y_{max}$) was defined as its arrival. The arrival time of each particle was recorded.

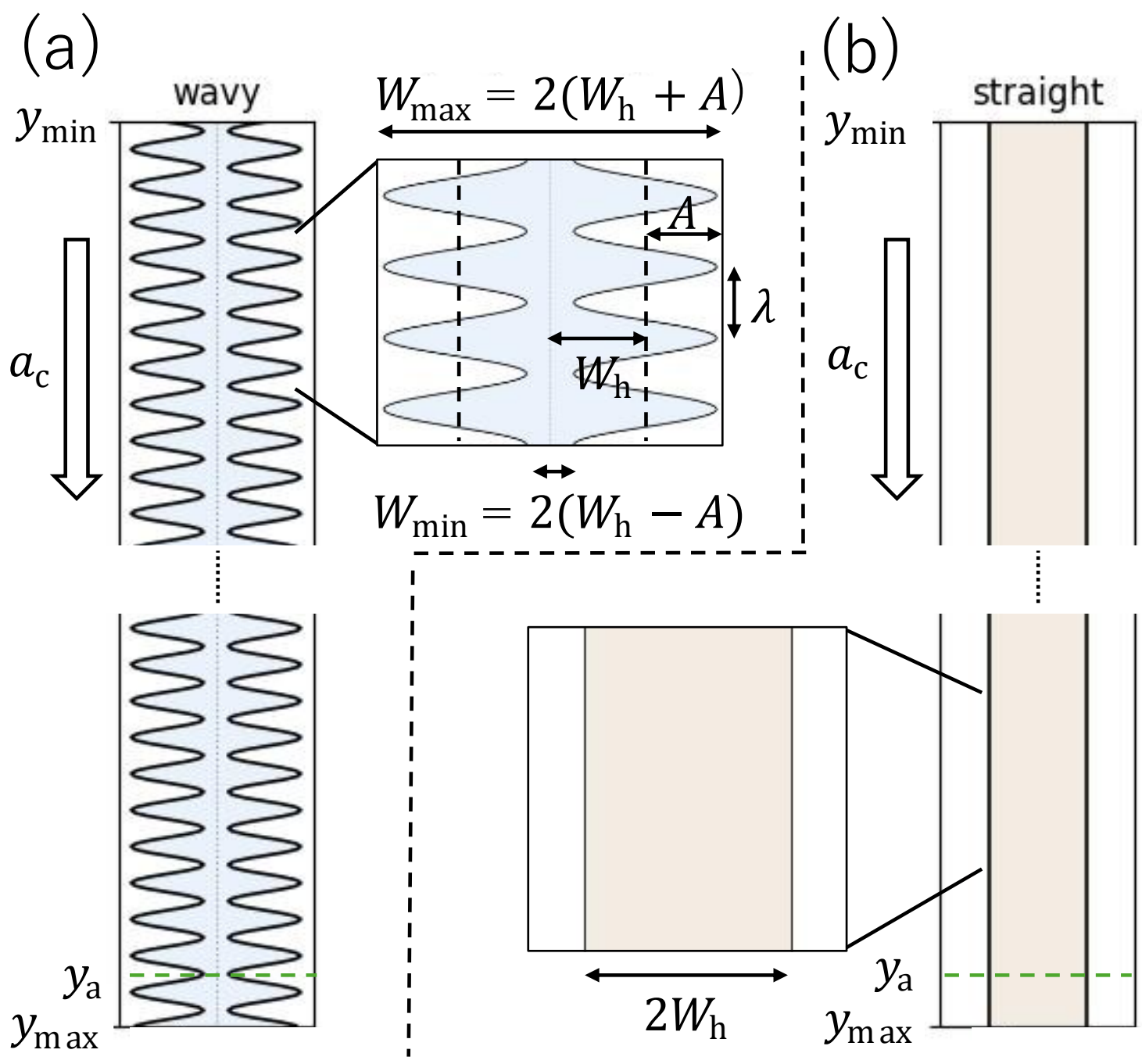

**Fig. 2.** Schematics of the channels: (a) wavy channel and (b) straight channel. The system was designed with the following values: $y_{min}$ = 0, $y_{max}$ = 10.1 mm, $y_a$ = 10 mm, $W_h$ = 20 μm, $A$ = 15 μm, and $\lambda$ = 15 μm. In (a), the maximum channel width $W_{max}$ is 70 μm, and the minimum width $W_{min}$ is 10 μm.

In the wavy channel, the periodic variation of the channel width changes the space available to particles depending on position, thereby forming geometric entropic barriers. The Fick–Jacobs approximation is applicable to dilute, overdamped systems in which particle–particle interactions can be neglected and the channel width varies sufficiently slowly along the transport direction (Malgaretti and Harting, 2023). The present wavy channel does not satisfy the latter condition, because its walls vary steeply, with an amplitude (15 μm) equal to the wavelength (15 μm). Therefore, Eqs. (8)–(10) below are used only for a qualitative interpretation of the entropic barriers, whereas the particle transport itself was computed directly by the two-dimensional Brownian dynamics simulations without this approximation. With the local channel width denoted by $W(y)$, the geometric entropy (Vazquez et al., 2008; Dagdug et al., 2012) is expressed as

$$S(y) = k_\mathrm{B} \ln W(y) \tag{8}$$

and the corresponding entropic potential $U$ is given by

$$U(y) = -TS(y) = -k_\mathrm{B} T \ln W(y). \tag{9}$$

Therefore, at the bottleneck, the entropy decreases, and the entropic potential accordingly increases, as the channel width decreases. In addition, because the effective diffusion coefficient in the Fick–Jacobs approximation also becomes position-dependent as the channel geometry changes, the diffusive transport of particles is suppressed in regions where the channel width changes abruptly (Zwanzig, 1992; Yariv and Dorfman, 2007). Furthermore, with the maximum channel width $W_\mathrm{max}$ and the minimum width at the bottlenecks $W_\mathrm{min}$, the height of the entropic barrier is expressed as

$$\Delta U = k_\mathrm{B} T \ln \left( \frac{W_\mathrm{max}}{W_\mathrm{min}} \right). \tag{10}$$

Thus, the smaller $W_\mathrm{min}$ is, that is, the narrower the bottleneck opening, the larger the entropic barrier becomes and the more difficult it is for particles to pass through. Because the response of particles to such entropic barriers depends on the relative contributions of centrifugal drift and Brownian diffusion, differences in the Péclet number arising from differences in particle size and density appear as differences in the passage behavior through the bottlenecks. Consequently, transport behavior differs even between particles with equal or similar sedimentation velocities, which is considered to enable temporal separation.

## 3. Results and discussion

### *3.1. Separation of particles with different sedimentation velocities by the centrifugal Brownian ratchet method*

In this section, the effectiveness of the centrifugal Brownian ratchet method for separating particles with different sedimentation velocities was examined under two conditions. The separation of particles with equal sedimentation velocities will be examined in Section 3.2.

In Condition 1, particles A1 and B1 both had a density of 1050 kg/m$^3$, and their diameters were 100 nm and 200 nm, respectively. The centrifugal acceleration was 9.807 × 10$^4$ m/s$^2$ (10 000 × $g$). The diffusion coefficients of particles A1 and B1 were $D_{\mathrm{A1}}$ = 4.37 × 10$^{-12}$ m$^2$/s and $D_{\mathrm{B1}}$ = 2.18 × 10$^{-12}$ m$^2$/s, and their sedimentation velocities were $v_{\mathrm{s,A1}}$ = 2.724 μm/s and $v_{\mathrm{s,B1}}$ = 10.896 μm/s, respectively; the corresponding Péclet numbers were $N_{\mathrm{P,A1}}$ = 16.0 and $N_{\mathrm{P,B1}}$ = 127.8. In Condition 2, particles A2 and B2 both had a diameter of 100 nm, with densities of 1012.5 kg/m$^3$ and 1050 kg/m$^3$, respectively, and the centrifugal acceleration was 6.865 × 10$^5$ m/s$^2$ (70 000 × $g$). In this case, particles A2 and B2 have the same diffusion coefficient of 4.37 × 10$^{-12}$ m$^2$/s but different sedimentation velocities, $v_{\mathrm{s,A2}}$ = 4.767 μm/s and $v_{\mathrm{s,B2}}$ = 19.068 μm/s, and therefore also different Péclet numbers, $N_{\mathrm{P,A2}}$ = 28.0 and $N_{\mathrm{P,B2}}$ = 111.8. All other parameters were the same as those described in Sections 2.1 and 2.2.

Fig. 3 shows the simulation results. To evaluate the separation performance quantitatively, the separation resolution $R_{\mathrm{s}}$ was used. The separation resolution $R_{\mathrm{s}}$ is an index that normalizes the difference between the mean positions of particles A and B by the combined spread of their respective distributions and is defined as

$$R_{\mathrm{s}} = \frac{|\mu_{\mathrm{B}} - \mu_{\mathrm{A}}|}{2(\sigma_{\mathrm{A}} + \sigma_{\mathrm{B}})}, \tag{11}$$

where $\mu_{\mathrm{A}}$ and $\mu_{\mathrm{B}}$ are the mean positions of particles A and B, respectively, and $\sigma_{\mathrm{A}}$ and $\sigma_{\mathrm{B}}$ are their respective standard deviations. A larger $R_{\mathrm{s}}$ indicates less overlap between the two distributions, and $R_{\mathrm{s}} > 1$ indicates that the two particle types are well separated.

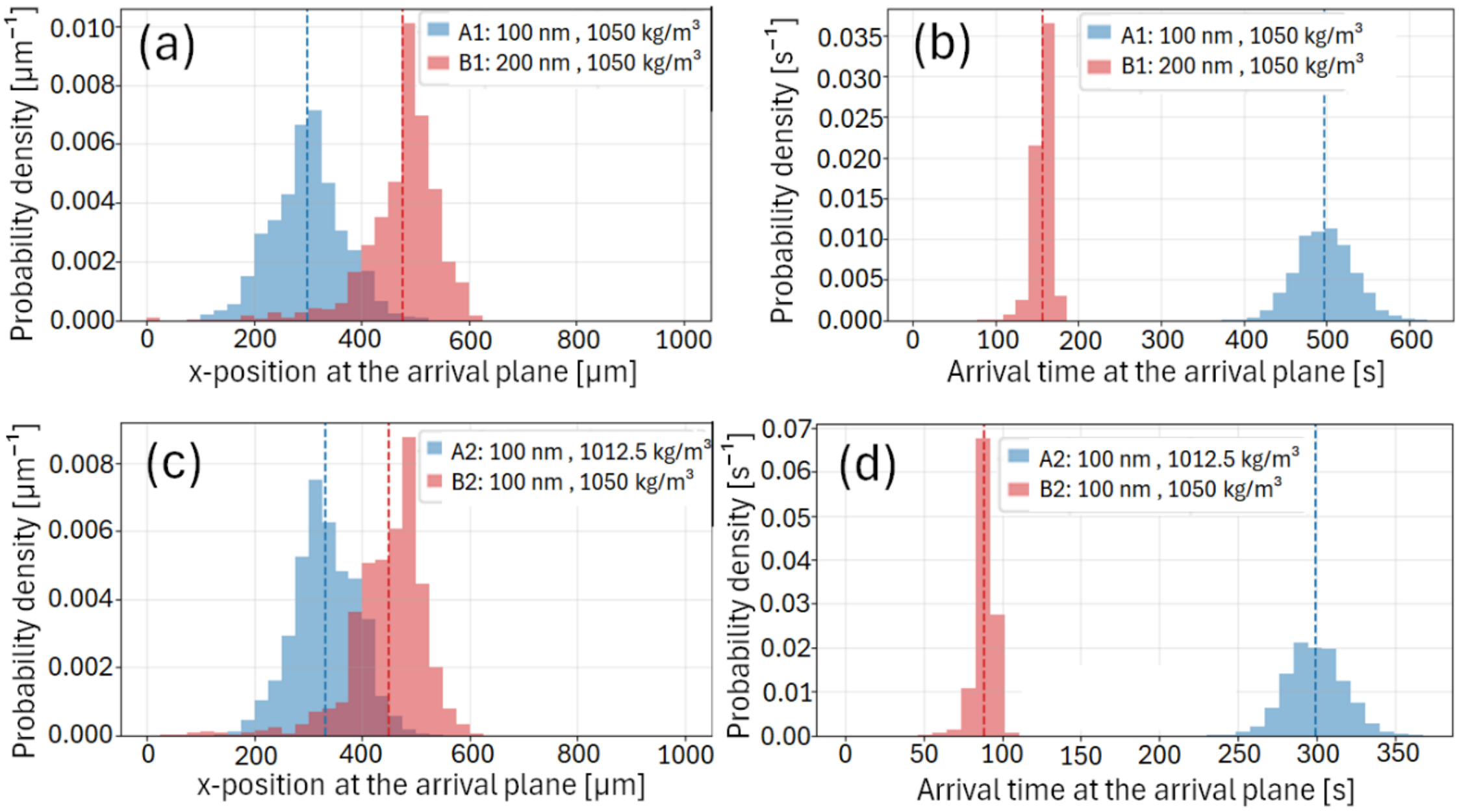

**Fig. 3.** Simulation results for the separation of particles with different sedimentation velocities by the centrifugal Brownian ratchet method. Position distributions at the arrival plane are shown in panels (a) and (c), while arrival-time distributions are shown in panels (b) and (d). Panels (a) and (b) correspond to Condition 1, and panels (c) and (d) correspond to Condition 2. Blue and red denote particles A and B, respectively. Dashed lines indicate the mean values.

First, the results for Condition 1 are described. As shown in Fig. 3(a), the mean $x$ positions at the arrival plane were 298.6 μm for particle A1 and 476.4 μm for particle B1, giving a difference in mean position of 177.7 μm. The separation resolution was $R_s$ = 0.65; because this did not exceed the criterion adopted in this study, lateral separation at the arrival plane could not be regarded as achieved. However, the width and length of the simulated centrifuge tube were both 1 mm, whereas centrifuge tubes several centimeters long are used in actual centrifugation experiments. Therefore, extending the width and length of the tube would increase the particle travel distance and is expected to improve separation performance.

For the arrival times shown in Fig. 3(b), the mean arrival times were 497.9 s for particle A1 and 156.8 s for particle B1, a difference of 341.1 s. The separation resolution for the arrival time was $R_s = 3.88$, indicating that the arrival-time distributions of the two particle types were clearly separated. Thus, Condition 1 demonstrated that the particles were clearly separable in arrival time.

Next, the results for Condition 2 are described. As shown in Fig. 3(c), the mean $x$ positions at the arrival plane were 332.0 μm for particle A2 and 449.9 μm for particle B2, giving a difference in mean position of 117.9 μm. The separation resolution was $R_s$ =

0.47; although the mean positions differed, the criterion for good separation was not reached. Thus, in Condition 2, the lateral transport properties differed, but sufficient separation was not achieved. However, in actual systems, the tube width and length are both larger than those used in the present simulations, which increases the particle travel distance and is expected to enable lateral separation in Condition 2 as well.

In contrast, for the arrival times shown in Fig. 3(d), the mean arrival times were 299.4 s for particle A2 and 88.6 s for particle B2, a difference of 210.8 s. The separation resolution for the arrival time was $R_s = 4.41$, indicating that the arrival-time distributions of the two particle types were completely separated without overlap. Thus, in Condition 2, although the lateral separation resolution was insufficient, a pronounced difference arose in the arrival time, showing that clear separation is possible in the time domain.

These results show that the centrifugal Brownian ratchet method produces differences in transport behavior between particles with different sedimentation velocities and that these differences can be detected as differences in arrival position and arrival time. In both Condition 1, in which particle sizes differed, and Condition 2, in which the particles had the same size and equal diffusion coefficients, differences in arrival position and arrival time arose from the difference in the Péclet number. These results suggest that the centrifugal Brownian ratchet method can serve as a particle separation technique that exploits differences in transport properties arising from the competition between the external force and Brownian diffusion, in addition to differences in sedimentation velocity.

### *3.2. Separation of particles with equal sedimentation velocities by the centrifugal Brownian ratchet method*

In Condition 3, the diameters of particles A3 and B3 were 100 nm and 200 nm, and their densities were 1050 kg/m$^3$ and 1012.5 kg/m$^3$, respectively. The centrifugal acceleration was $3.923 \times 10^5$ m/s$^2$ (40 000 × $g$). All other conditions were the same as those described in Sections 2.1 and 2.2. Although these particles differ in size and density, both have the same sedimentation velocity of 10.896 µm/s. In a system without ratchets, no clear difference arose between particles A3 and B3 in the mean arrival position or the mean arrival time at the arrival plane. However, the diffusion coefficients and Péclet numbers were $D_{\mathrm{A3}} = 4.37 \times 10^{-12}$ m$^2$/s and $N_{\mathrm{P,A3}} = 63.9$ for particle A3 and $D_{\mathrm{B3}} = 2.18 \times 10^{-12}$ m$^2$/s and $N_{\mathrm{P,B3}} = 127.8$ for particle B3, so the relative contributions of Brownian diffusion and externally driven transport differ between the two particle types. We therefore examined whether particles with equal sedimentation velocities can be separated by the centrifugal Brownian ratchet method.

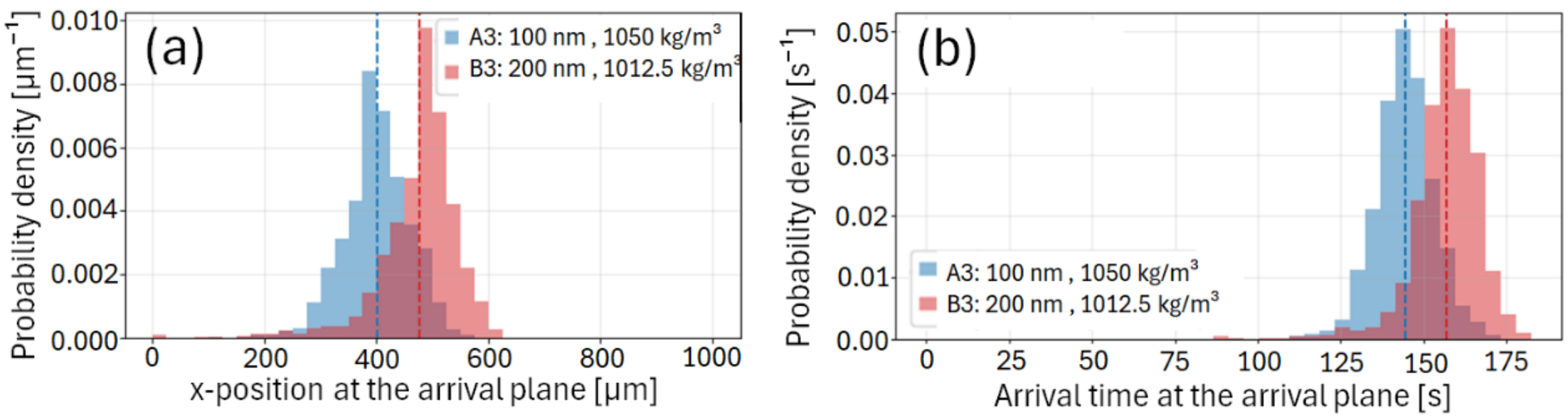

**Fig. 4.** Simulation results for the separation of particles with equal sedimentation velocities by the centrifugal Brownian ratchet method (Condition 3): (a) position distributions and (b) arrival-time distributions at the arrival plane. Blue and red denote particles A and B, respectively. Dashed lines indicate mean values.

As shown in Fig. 4(a), the mean $x$ positions at the arrival plane were 400.3 μm for particle A3 and 476.6 μm for particle B3, giving a difference in mean position of 76.3 μm. The separation resolution was $R_s$ = 0.30, which did not reach the criterion for good separation, $R_s$ > 1, described above.

For the arrival-time distributions shown in Fig. 4(b), the mean arrival times were 144.4 s for particle A3 and 156.8 s for particle B3, a difference of 12.4 s. The separation resolution for the arrival time was $R_s$ = 0.33, which also did not reach the criterion of $R_s$ > 1. Thus, although the transport behaviors of particles A3 and B3 also differed clearly in arrival time, the distributions could not be regarded as completely separated. However, in actual systems, the tube width and length are both larger than those used in the present simulations, which increases the particle travel distance and is therefore expected to enable both lateral and temporal separation in Condition 3.

It is noteworthy that, although particles A3 and B3 had equal sedimentation velocities, their mean arrival times differed by 12.4 s, with particle A3 reaching the arrival plane in a shorter time than particle B3. In Conditions 1 and 2, particles with larger Péclet numbers reached the arrival plane in a shorter time; however, these particles also had larger sedimentation velocities. When the mean arrival time is multiplied by $v_s$ and divided by the sedimentation distance $y_a$, the resulting normalized arrival time increases monotonically with the Péclet number in all three conditions: 1.36 (A1, $N_P$ = 16.0), 1.43 (A2, 28.0), 1.57 (A3, 63.9), 1.69 (B2, 111.8), 1.71 (B1, 127.8), and 1.71 (B3, 127.8). The

longer arrival time of particle B3 is therefore consistent with this trend rather than an exception. A possible reason is that the larger Péclet number of particle B3 enhanced centrifugal-force-driven transport relative to Brownian diffusion, resulting in stronger lateral rectification by the ratchet structures. Thus, despite identical sedimentation velocities, the difference in Péclet number may have altered the interaction with the ratchets and the effective transport path, leading to the difference in arrival time. This property suggests that particle properties might be estimated by comparing where and when the particles arrive.

### *3.3. Separation of particles with different sedimentation velocities by the centrifugal wavy-channel method*

Next, the effectiveness of the centrifugal wavy-channel method for separating particles with different sedimentation velocities was examined. The separation of particles with equal sedimentation velocities will be examined in Section 3.4. In Conditions 4 and 5, the particle sizes, mass densities, and centrifugal accelerations were set identical to those in Conditions 1 and 2, respectively. The diffusion coefficients and sedimentation velocities are therefore identical to those in Conditions 1 and 2, respectively. The modified Péclet numbers of particles A4 and B4 in Condition 4 were $N_{\mathrm{P,w,A4}} = 33.26$ and $N_{\mathrm{P,w,B4}} = 266.11$, respectively, while those of particles A5 and B5 in Condition 5 were $N_{\mathrm{P,w,A5}} = 58.21$ and $N_{\mathrm{P,w,B5}} = 232.85$, respectively. All other conditions were the same as those described in Sections 2.1 and 2.3.

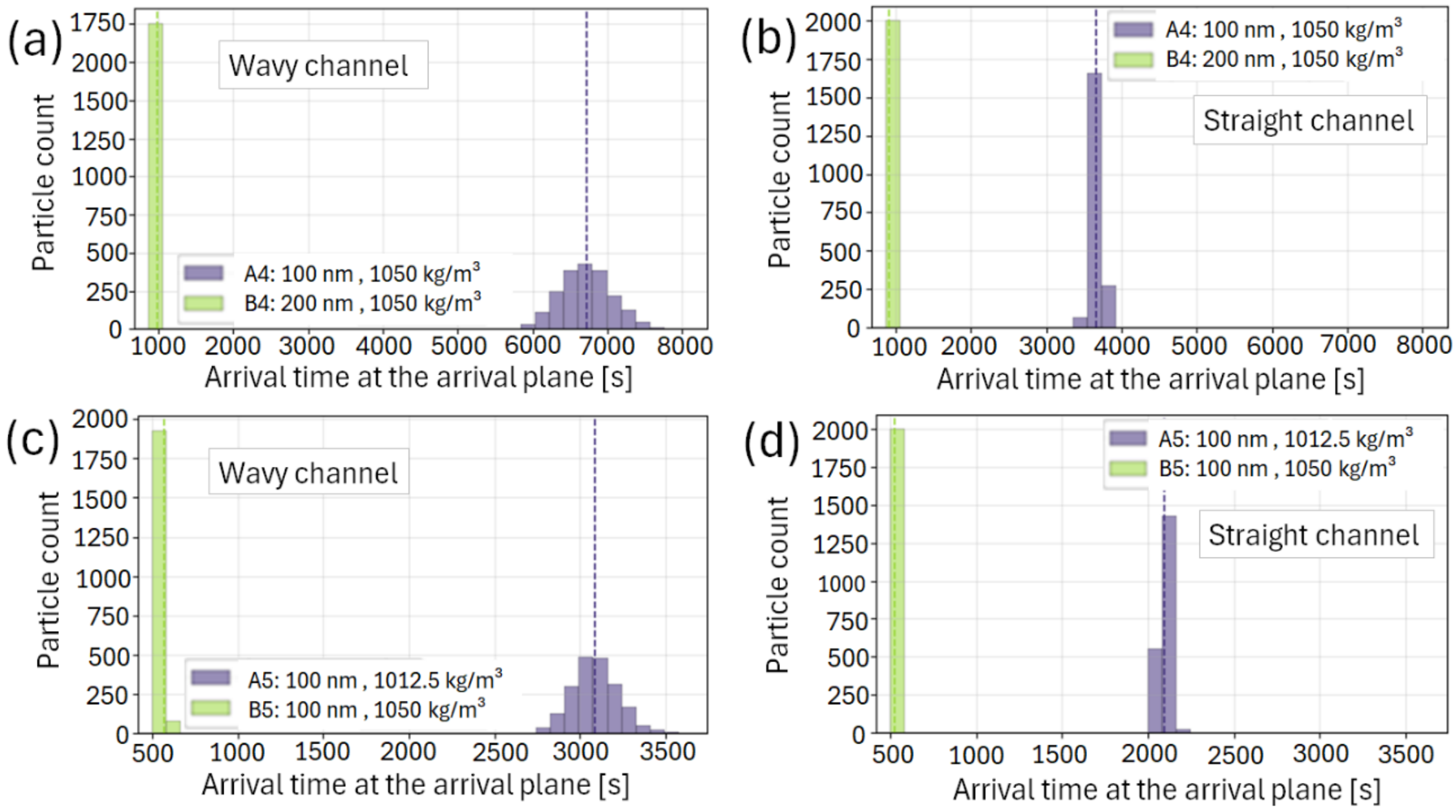

**Fig. 5.** Simulation results for the separation of particles with different sedimentation velocities by the centrifugal wavy-channel method. Arrival-time distributions at the arrival plane are shown for the wavy and straight channels. Panels (a) and (b) correspond to Condition 4, while panels (c) and (d) correspond to Condition 5. Panels (a) and (c) show the results for the wavy channel, whereas panels (b) and (d) show those for the straight channel. Purple and green denote particles A and B, respectively. Dashed lines indicate the mean values.

Fig. 5 compares the arrival times at the arrival plane in the wavy and straight channels for Conditions 4 and 5. First, the results for Condition 4 are described. As shown in Fig. 5(a), in the wavy channel, particles A4 and B4 arrived at mean times of 6718.3 s and 986.5 s, respectively, giving a difference of 5731.8 s and a separation resolution of $R_s =$ 8.05. In the straight channel, by contrast, particles A4 and B4 arrived at mean times of 3667.3 s and 917.2 s, respectively, giving a difference of 2750.1 s and a separation resolution of $R_s = 19.31$, as shown in Fig. 5(b). In both channels, $R_s$ far exceeded the criterion for sufficient separation ($R_s > 1$). Although $R_s$ was higher in the straight channel owing to suppressed peak broadening, the difference in mean arrival time was 2981.7 s larger in the wavy channel, demonstrating that the wavy channel substantially expands the temporal separation window.

Next, the results for Condition 5 are described. As shown in Fig. 5(c), in the wavy channel, particles A5 and B5 arrived at mean times of 3088.9 s and 570.1 s, respectively, giving a difference of 2518.8 s and a separation resolution of $R_s = 8.96$. In the straight channel, by contrast, particles A5 and B5 arrived at mean times of 2096.6 s and 524.0 s, respectively, giving a difference of 1572.6 s and a separation resolution of $R_s = 24.65$, as shown in Fig. 5(d). In both channels, $R_s$ far exceeded the criterion for sufficient separation ($R_s > 1$). Although $R_s$ was higher in the straight channel, the arrival-time difference between particles A5 and B5 was 946.2 s larger in the wavy channel, confirming that the wavy channel is advantageous in Condition 5 as well by widening the temporal separation window. These results demonstrate that using a wavy channel effectively enhances the arrival-time difference between particles with different sedimentation velocities.

### *3.4. Separation of particles with equal sedimentation velocities by the centrifugal wavy-channel method*

We examined whether particles with equal sedimentation velocities can be separated by the centrifugal wavy-channel method. In Condition 6, the particle sizes, mass densities, and centrifugal acceleration were set identical to those in Condition 3. The diffusion coefficients and sedimentation velocities are therefore identical to those in Condition 3. The modified Péclet numbers of particles A6 and B6 in Condition 6 were $N_{P,w,A6} = 133.05$ and $N_{P,w,B6} = 266.11$, respectively. All other conditions were the same as those described in Sections 2.1 and 2.3. Because the sedimentation velocities of the two particle types are equal, the mean arrival times of particles A6 and B6 at the arrival plane in the straight channel were 917.4 s and 917.2 s, respectively, with almost no difference and a separation resolution of $R_s = 0.01$.

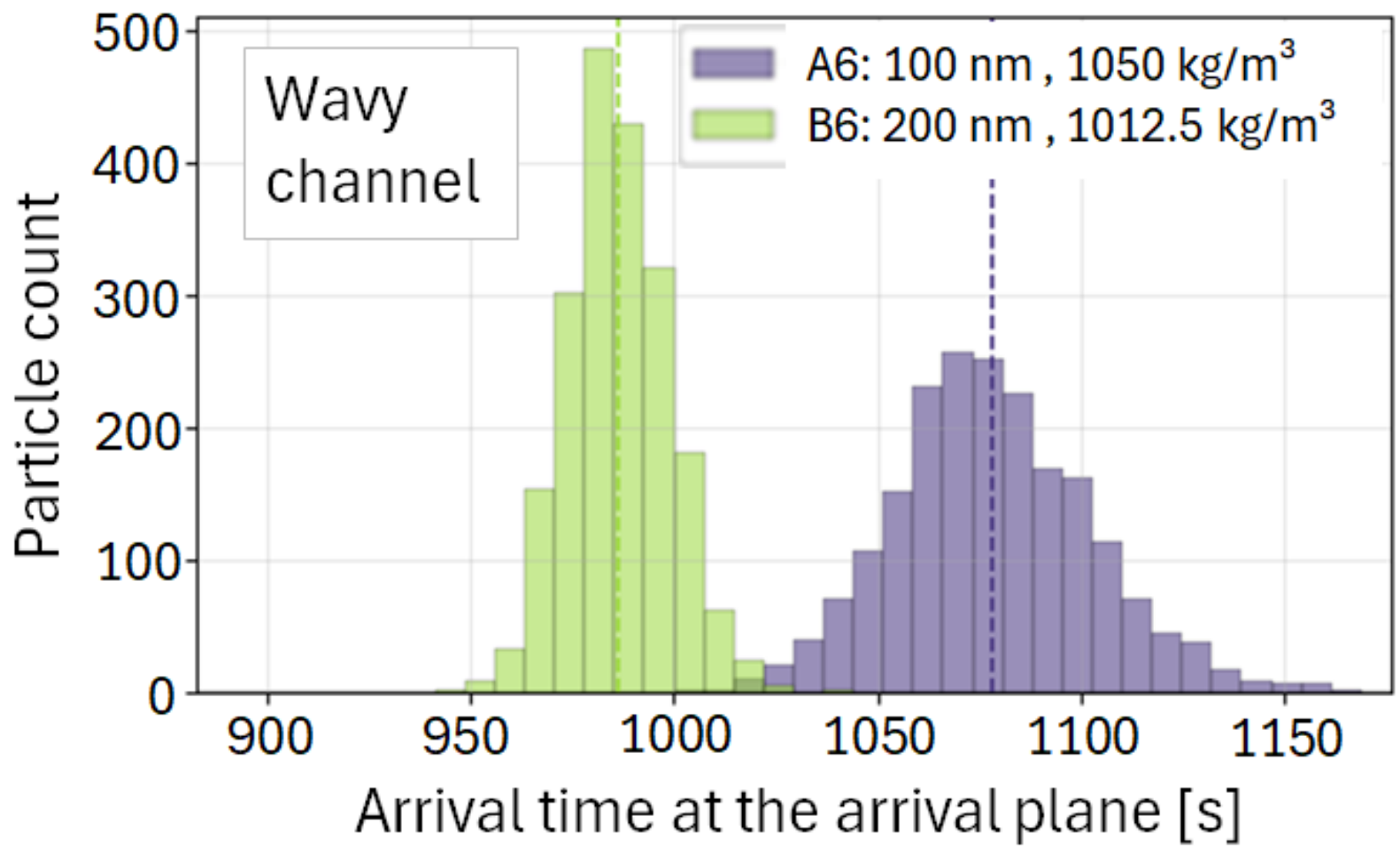

**Fig. 6.** Simulation results for the separation of particles with equal sedimentation velocities by the centrifugal wavy-channel method: arrival-time distributions at the arrival plane in the wavy channel for Condition 6. Purple and green denote particles A and B, respectively. Dashed lines indicate mean values.

Fig. 6 shows the arrival-time distributions at the arrival plane for particles A6 and B6 in the wavy channel. As shown in Fig. 6, in the wavy channel, particles A6 and B6 arrived at mean times of 1078.3 s and 986.4 s, respectively, giving a difference of 91.9 s and a separation resolution of $R_\mathrm{s} = 1.26$. Thus, even for particles with equal sedimentation velocities, for which virtually no arrival-time difference arose in the straight channel ($R_\mathrm{s} = 0.01$), the wavy channel produced a distinct difference in arrival time and substantially enhanced separation capability.

In addition, a simulation was performed in which $W_\mathrm{max}$ was changed from 70 μm to 90 μm while $W_\mathrm{min}$ and all other parameters were kept unchanged, so that a larger entropic barrier was imposed. As a result, the mean arrival times of particles A6 and B6 were 1117.7 s and 1002.0 s, respectively, yielding an increased arrival-time difference of 115.7 s and a separation resolution of $R_\mathrm{s} = 1.25$. These results show that applying a wavy channel with a higher entropic barrier can further widen the arrival-time difference between particles, confirming that the magnitude of the temporal separation window expands with increasing entropic barrier height.

*3.5. Can particles with equal Péclet numbers be separated?*

In Condition 7, the diameters of particles A7 and B7 were 100 nm and 200 nm, and their densities were 1050 kg/m$^3$ and 1006.25 kg/m$^3$, respectively. The centrifugal acceleration was 4.903 × 10$^5$ m/s$^2$ (50 000 × $g$). In this case, the diffusion coefficients are $D_\mathrm{A7}$ = 4.37 × 10$^{-12}$ m$^2$/s and $D_\mathrm{B7}$ = 2.18 × 10$^{-12}$ m$^2$/s, the sedimentation velocities are $v_\mathrm{s,A7}$ = 13.620 μm/s and $v_\mathrm{s,B7}$ = 6.810 μm/s, and both Péclet numbers are 79.9.

First, separation simulations were performed with the centrifugal Brownian ratchet method. Apart from the settings of Condition 7, all conditions were the same as those described in Sections 2.1 and 2.2. The mean $x$ positions at the arrival plane were 419.6 μm for particle A7 and 433.8 μm for particle B7, giving a difference in mean position of 14.2 μm. The separation resolution was $R_\mathrm{s} = 0.055$, showing that the difference between the distributions of the two particle types was relatively small. In contrast, the mean arrival times were 118.6 s for particle A7 and 237.1 s for particle B7, giving a difference of 118.5 s and a separation resolution of $R_\mathrm{s} = 2.70$, which well exceeds the criterion for sufficient separation ($R_\mathrm{s} > 1$). These results reveal that, for particles with equal Péclet numbers, the centrifugal Brownian ratchet method enables separation based on arrival time, whereas lateral separation is difficult. Lateral spatial separation in this method is therefore considered to arise primarily from differences in the Péclet number between particles. Nevertheless, because the method enables lateral or temporal separation of particles with equal or similar sedimentation velocities, which are difficult to separate by conventional centrifugal sedimentation, it can still be a useful separation technique.

Next, separation simulations were performed under the same conditions with the centrifugal wavy-channel method. In Condition 8, the particle sizes, mass densities, and centrifugal acceleration were set identical to those in Condition 7. The diffusion coefficients and sedimentation velocities are therefore identical to those in Condition 7. The modified Péclet numbers of particles A8 and B8 in Condition 8 were both 166.32. All other conditions were the same as those described in Sections 2.1 and 2.3. In the wavy channel, particles A8 and B8 arrived at mean times of 831.7 s and 1663.9 s, respectively, giving a difference of 832.2 s and a separation resolution of $R_s = 9.02$. In the straight channel, by contrast, particles A8 and B8 arrived at mean times of 733.8 s and 1467.5 s, respectively, giving a difference of 733.7 s and a separation resolution of $R_s = 20.85$. In both channels, $R_s$ far exceeded the criterion for sufficient separation ($R_s > 1$). Although $R_s$ was higher in the straight channel, applying the wavy channel increased the mean arrival-time difference by 98.5 s, thereby further widening the temporal separation window. These results show that, in the centrifugal wavy-channel method, arrival-time differences can be generated and enlarged through the combined effects of the wavy-channel structure and differences in diffusion coefficients, achieving effective separation even for particles with identical Péclet numbers.

Finally, several simplifications were adopted in the present simulations to isolate the fundamental separation mechanisms: a two-dimensional geometry, monodisperse particles, a spatially uniform centrifugal field in a small computational domain, and the neglect of hydrodynamic interactions between particles and with the walls. These simplifications are not expected to alter the qualitative mechanisms identified here; however, quantitative predictions for practical devices will require three-dimensional simulations that include hydrodynamic interactions, particle polydispersity, and the radial variation of the centrifugal field, together with experimental validation using actual particles such as DDS nanoparticles and exosomes.

## 4. Conclusions

In this study, the centrifugal Brownian ratchet method and the centrifugal wavy-channel method were proposed as methods for separating small particles with equal or similar sedimentation velocities, and the separation performance and transport mechanisms of the two methods were compared and evaluated through Brownian dynamics simulations. Building on these results, we envision a two-step separation process as a practical application: particles are first separated by sedimentation velocity using conventional centrifugation, with particles having equal sedimentation velocities collected into the same band, and these particles are then further separated using either the centrifugal Brownian ratchet or centrifugal wavy-channel method based on differences in Péclet number and transport properties.

In addition to these separation capabilities, neither method requires osmotic stress from high-concentration media or the imparting of electrical properties, and both are therefore expected to separate particles in diverse solvents, including liquid metals, while preserving their properties and surface characteristics. The centrifugal Brownian ratchet method enables spatial separation without strict time control, whereas the centrifugal wavy-channel method enables faster recovery of particles with relatively high sedimentation velocities under comparable tube depths and centrifugal accelerations. Owing to its simple channel geometry, the wavy-channel method is considered to be more readily applicable to future separation mechanisms that exploit differences in particle charge.

**Table 1.** Comparison of the features of centrifugal separation methods.

| Method | Separation | Principle | Separability for similar $v_s$ | Separability for similar $N_P$ | Particle Damage |
|---|---|---|---|---|---|
| Brownian Ratchet | L(i) | Péclet number | ○ | ×(ii) | Low |
| Wavy Channel | V | Entropic Barrier | ○ | ○(iii) | Low |
| Conventional Sedimentation | V | Sedimentation velocity | × | ○ | Low |
| Conventional Density Gradient | S | Buoyant density | ○ | ○ | Middle(iv) |

L: Lateral ($x$-direction) positional separation

V: Vertical ($y$-direction) temporal/spatial separation

S: Vertical ($y$ -direction) spatial separation

(i): Temporal separation is possible even when the lateral separation is not.

(ii): Lateral separation is difficult, but the temporal separation is possible.

(iii): It is advantageous compared with conventional centrifugal sedimentation because it increases the difference in particle arrival times.

(iv): For delicate particles, such as DDS nanoparticles and exosomes, high osmotic pressure may compromise their structures and/or surface properties.

As summarized in Table 1, the proposed centrifugal Brownian ratchet and wavy-channel methods are likely to offer advantages over conventional centrifugal sedimentation and density gradient centrifugation, particularly for the separation of DDS nanoparticles and exosomes. First, density gradient centrifugation requires a medium with a high solute concentration, and the resulting osmotic pressure may cause these delicate particles to swell, shrink, or rupture (Konoshenko et al., 2018). The proposed methods require no such medium and therefore avoid this risk. Second, samples of DDS nanoparticles and exosomes can contain particles whose different combinations of size and density result in similar sedimentation velocities, so that conventional centrifugal sedimentation, which relies solely on differences in sedimentation velocity, provides limited resolution for these particles. In contrast, the centrifugal wavy-channel method separates particles along the sedimentation ($y$) direction, that is, by arrival time, and enlarges arrival-time differences through periodic entropic barriers even for particles with equal sedimentation velocities (Section 3.4). Likewise, the centrifugal Brownian ratchet method produces differences in both lateral position and arrival time between particles with equal sedimentation velocities by exploiting differences in the Péclet number (Section 3.2), and these differences are expected to become resolvable with longer centrifuge tubes. These features make both methods promising candidates for the gentle, high-resolution separation of delicate particles such as DDS nanoparticles and exosomes, although experimental validation with such particles remains to be performed.

### CRediT authorship contribution statement

**Yurika Miyake:** Writing – review & editing, Writing – original draft, Visualization, Validation, Software, Methodology, Investigation, Data curation. **Natsuki Asami:** Writing – review & editing, Visualization, Validation, Investigation. **Ken-ichi Amano:** Writing – review & editing, Writing – original draft, Visualization, Validation, Supervision, Software, Resources, Project administration, Methodology, Investigation, Formal analysis, Data curation, Conceptualization.

### Declaration of competing interest

The authors declare that they have no known competing financial interests or personal relationships that could have appeared to influence the work reported in this paper.

### Data availability

Data will be made available on request.

### Declaration of generative AI and AI-assisted technologies in the manuscript preparation process

During the preparation of this work, the Japanese manuscript was written entirely by the authors, and AI-assisted tools were used to translate the manuscript from Japanese into English. The basic design and structure of the computational programs were independently developed by the authors; however, AI-assisted tools were used to help revise and modify the code. After using these tools, the authors reviewed and edited the AI-generated content as needed and take full responsibility for the content of the published article.

## Acknowledgements

We thank M. Maebayashi for providing a supportive research environment.

## Funding

This research did not receive any specific grant from funding agencies in the public, commercial, or not-for-profit sectors.

## References

Astumian, R.D., 1997. Thermodynamics and kinetics of a Brownian motor. Science 276, 917–922. https://doi.org/10.1126/science.276.5314.917.

Berezhkovskii, A.M., Dagdug, L., Bezrukov, S.M., 2015. Range of applicability of modified Fick–Jacobs equation in two dimensions. J. Chem. Phys. 143, 164102. https://doi.org/10.1063/1.4934223.

Burada, P.S., Schmid, G., Reguera, D., Rubí, J.M., Hänggi, P., 2007. Biased diffusion in confined media: test of the Fick–Jacobs approximation and validity criteria. Phys. Rev. E 75, 051111. https://doi.org/10.1103/PhysRevE.75.051111.

Burada, P.S., Schmid, G., Talkner, P., Hänggi, P., Reguera, D., Rubí, J.M., 2008. Entropic particle transport in periodic channels. BioSystems 93, 16–22. https://doi.org/10.1016/j.biosystems.2008.03.006.

Dagdug, L., Vazquez, M.-V., Berezhkovskii, A.M., Zitserman, V.Y., Bezrukov, S.M., 2012. Diffusion in the presence of cylindrical obstacles arranged in a square lattice analyzed with generalized Fick–Jacobs equation. J. Chem. Phys. 136, 204106. https://doi.org/10.1063/1.4720385.

Devendra, R., Drazer, G., 2012. Gravity driven deterministic lateral displacement for particle separation in microfluidic devices. Anal. Chem. 84, 10621–10627. https://doi.org/10.1021/ac302074b.

Dixit, P.D., Dill, K.A., 2014. Inferring microscopic kinetic rates from stationary state distributions. J. Chem. Theory Comput. 10, 3002–3005. https://doi.org/10.1021/ct5001389.

Du, S., Drazer, G., 2016. Gravity driven deterministic lateral displacement for suspended particles in a 3D obstacle array. Sci. Rep. 6, 31428. https://doi.org/10.1038/srep31428.

Grimm, A., Stark, H., van der Maarel, J.R.C., 2009. Model for a Brownian ratchet with improved characteristics for particle separation. Phys. Rev. E 79, 061102. https://doi.org/10.1103/PhysRevE.79.061102.

Hänggi, P., Marchesoni, F., Nori, F., 2005. Brownian motors. Ann. Phys. 517, 51–70.

https://doi.org/10.1002/andp.200551701-304.

Huang, L.R., Cox, E.C., Austin, R.H., Sturm, J.C., 2003. Tilted Brownian ratchet for DNA analysis. Anal. Chem. 75, 6963–6967. https://doi.org/10.1021/ac0348524.

Jiang, M., Mazzeo, A.D., Drazer, G., 2016. Centrifuge-based deterministic lateral displacement separation. Microfluid. Nanofluid. 20, 17. https://doi.org/10.1007/s10404-015-1686-x.

Kalinay, P., Percus, J.K., 2006. Corrections to the Fick–Jacobs equation. Phys. Rev. E 74, 041203. https://doi.org/10.1103/PhysRevE.74.041203.

Konoshenko, M.Y., Lekchnov, E.A., Vlassov, A.V., Laktionov, P.P., 2018. Isolation of extracellular vesicles: general methodologies and latest trends. BioMed Res. Int. 2018, 8545347. https://doi.org/10.1155/2018/8545347.

Kowalik, M., Bishop, K.J.M., 2016. Ratcheted electrophoresis of Brownian particles. Appl. Phys. Lett. 108, 203103. https://doi.org/10.1063/1.4950801.

Laachi, N., Kenward, M., Yariv, E., Dorfman, K.D., 2007. Force-driven transport through periodic entropy barriers. EPL 80, 50009. https://doi.org/10.1209/0295-5075/80/50009.

Linke, H., Downton, M.T., Zuckermann, M.J., 2005. Performance characteristics of Brownian motors. Chaos 15, 026111. https://doi.org/10.1063/1.1871432.

Malgaretti, P., Harting, J., 2023. Closed formula for transport across constrictions. Entropy 25, 470. https://doi.org/10.3390/e25030470.

Norouzi, N., Bhakta, H.C., Grover, W.H., 2017. Sorting cells by their density. PLoS One 12, e0180520. https://doi.org/10.1371/journal.pone.0180520.

Plüisch, C.S., Bössenecker, B., Dobler, L., Wittemann, A., 2019. Zonal rotor centrifugation revisited: new horizons in sorting nanoparticles. RSC Adv. 9, 27549–27559. https://doi.org/10.1039/c9ra05140f.

Reguera, D., Rubí, J.M., 2001. Kinetic equations for diffusion in the presence of entropic barriers. Phys. Rev. E 64, 061106. https://doi.org/10.1103/PhysRevE.64.061106.

Reguera, D., Schmid, G., Burada, P.S., Rubí, J.M., Reimann, P., Hänggi, P., 2006. Entropic transport: kinetics, scaling, and control mechanisms. Phys. Rev. Lett. 96, 130603. https://doi.org/10.1103/PhysRevLett.96.130603.

Reguera, D., Luque, A., Burada, P.S., Schmid, G., Rubí, J.M., Hänggi, P., 2012. Entropic splitter for particle separation. Phys. Rev. Lett. 108, 020604. https://doi.org/10.1103/PhysRevLett.108.020604.

Reimann, P., 2002. Brownian motors: noisy transport far from equilibrium. Phys. Rep. 361, 57–265. https://doi.org/10.1016/S0370-1573(01)00081-3.

Riefler, W., Schmid, G., Burada, P.S., Hänggi, P., 2010. Entropic transport of finite size particles. J. Phys.: Condens. Matter 22, 454109. https://doi.org/10.1088/0953-8984/22/45/454109.

Schmid, G., Burada, P.S., Talkner, P., Hänggi, P., 2009. Rectification through entropic barriers. Adv. Solid State Phys. 48, 317–328. https://doi.org/10.1007/978-3-540-85859-1_25.

Smart, J., Lattuada, M., 2026. Separation of colloidal particles by density gradient centrifugation: techniques and practical guidelines. Adv. Colloid Interface Sci. 357, 103993. https://doi.org/10.1016/j.cis.2026.103993.

Vazquez, M.-V., Berezhkovskii, A.M., Dagdug, L., 2008. Diffusion in linear porous media with periodic entropy barriers: a tube formed by contacting spheres. J. Chem. Phys. 129, 046101. https://doi.org/10.1063/1.2955447.

Xuan, J., Lee, M.L., 2014. Size separation of biomolecules and bioparticles using micro/nanofabricated structures. Anal. Methods 6, 27–37. https://doi.org/10.1039/c3ay41364k.

Yakubovich, E.I., Polischouk, A.G., Evtushenko, V.I., 2022. Principles and problems of exosome isolation from biological fluids. Biochem. (Mosc.) Suppl. Ser. A: Membr. Cell Biol. 16, 115–126. https://doi.org/10.1134/S1990747822030096.

Yang, D., Zhang, W., Zhang, H., Zhang, F., Chen, L., Ma, L., Larcher, L.M., Chen, S., Liu, N., Zhao, Q., Tran, P.H.L., Chen, C., Veedu, R.N., Wang, T., 2020. Progress, opportunity, and perspective on exosome isolation – efforts for efficient exosome-based theranostics. Theranostics 10, 3684–3707. https://doi.org/10.7150/thno.41580.

Yariv, E., Dorfman, K.D., 2007. Electrophoretic transport through channels of periodically varying cross section. Phys. Fluids 19, 037101. https://doi.org/10.1063/1.2710894.

Zhang, X.-J., Qian, H., Qian, M., 2012. Stochastic theory of nonequilibrium steady states and its applications. Part I. Phys. Rep. 510, 1–86. https://doi.org/10.1016/j.physrep.2011.09.002.

Zwanzig, R., 1992. Diffusion past an entropy barrier. J. Phys. Chem. 96, 3926–3930. https://doi.org/10.1021/j100189a004.